\documentstyle[pre,aps,eqsecnum]{revtex}
\begin{document}
\title{
Limited Range Fractality of Randomly Adsorbed Rods
}
\author{Daniel A. Lidar (Hamburger)$^{a,b}$\footnote{URL:
http://www.fh.huji.ac.il/$\!\sim$dani},
        Ofer Biham$^{a}$\footnote{URL:
http://www.fiz.huji.ac.il/staff/acc/faculty/biham} and 
        David Avnir$^{b,c,d}$\footnote{URL:
http://chem.ch.huji.ac.il/employee/avnir/iavnir.htm}}
\address{
$^{a}$ Racah Institute of Physics, The Hebrew University, Jerusalem
91904, Israel }
\address{
$^{b}$ Fritz Haber Center for Molecular Dynamics, The Hebrew University,
Jerusalem 91904, Israel }
\address{
$^{c}$ Institute of Chemistry, The Hebrew University, Jerusalem 91904, Israel
}
\address{
$^{d}$ Minerva Center for Computational Quantum Chemistry, 
The Hebrew University, Jerusalem 91904, Israel }
\maketitle

\begin{abstract}
\newline{}
Multiple resolution analysis of two dimensional 
structures composed of randomly adsorbed
penetrable rods, 
for densities below the percolation threshold,
has been carried out using
box-counting functions.
It is found that at relevant resolutions, 
for box-sizes, $r$,
between cutoffs given by 
the average rod length 
$\langle \ell \rangle$
and the average
inter-rod distance 
$r_1$,
these systems exhibit apparent
fractal behavior.
It is shown that unlike the case of randomly distributed isotropic
objects,
the upper cutoff 
$r_1$
is not only
a function of the coverage but also depends on the excluded volume,
averaged over the orientational distribution.
Moreover, the apparent fractal dimension also depends on the
orientational distributions of the rods and decreases as it 
becomes more anisotropic.
For box sizes
smaller than 
$\langle \ell \rangle$
the box
counting function is determined by the internal structure of the 
rods, whether simple or itself fractal. 
Two examples are considered -- one of regular rods
of one dimensional structure and rods which are trimmed
into a Cantor set structure which are fractals themselves.
The models examined are relevant to  
adsorption of linear molecules and fibers, 
liquid crystals, 
stress induced fractures and 
edge imperfections in metal catalysts.
We thus obtain a
distinction between two ranges of length scales:
$r < \langle \ell \rangle$
where the internal structure of the adsorbed objects is
probed, and
$\langle \ell \rangle < r < r_1$ where their distribution
is probed, both of which may exhibit fractal behavior.
This distinction is
relevant to the large
class of systems which exhibit aggregation of a finite 
density of fractal-like clusters,
which includes surface growth in molecular beam epitaxy
and diffusion-limited-cluster-cluster-aggregation models.
\end{abstract}



\section{Introduction}

The structure, packing and phase transitions in media
composed of rod-like particles are of great importance 
in a large variety of physical systems
\cite{Onsager2,Flory,Chick,viot92,ricci92}.
Such systems include liquid crystals
\cite{degennes74},
slip lines in stressed metals and rocks, 
line-fractures in catalysts
\cite{jost88,romeu86,mandelbrot84},
fibers in paper sheets
\cite{provatas95,provatas96},
and rod-like metal particles
\cite{zhang92,brunner95}.
Unlike systems composed of nearly isotropic particles,
for rod-like particles the orientational distribution plays an
important role 
in addition to the positional distribution
\cite{Balberg83}.
Systems of adsorbed particles can be classified to 
ones of penetrable 
\cite{torquato83,Pike74,Balberg84b}
vs. impenetrable particles
\cite{viot92,ricci92,Torquato:2,Evans93}. 
In this paper we will focus on the case of penetrable rods.

Many of the systems of interest in this context are
essentially two dimensional and can be studied by 
a model
in which $M$ penetrable rods are randomly distributed
on a two dimensional surface of size $L \times L$.
Each rod is of length
$\ell$ and width $w$. 
The effective two dimensional coverage 
is given by
$\eta_2 = \rho \cdot \ell  w$,
where
$\rho=M/L^2$
is the rod density.
It 
provides a measure of the fraction of the area which is covered
by the rods, where the reduction due to overlaps is ignored.
Clusters of overlapping rods start to form at low rod density 
and increase in size as the density increases.
Systems 
of this type
exhibit a percolation transition at rod density $\rho_c$, 
above which there is
an cluster which spans across the entire system.
The percolation transition for systems of rod-like particles 
has been studied extensively in recent years
\cite{Balberg83,Pike74,Balberg84b,Balberg84a,Bug85,Balberg87,Alon90,Alon91}. 
A criterion for the
percolation density $\rho_c$
was obtained 
and examined for a variety of systems with
different distributions of rod sizes and orientations.
According to this criterion, 
percolation for isotropically oriented and
randomly distributed rods occurs at
rod density 
$\rho_c \cong 1/A_{exc}$
\cite{Balberg84b}.
Here $A_{exc}$ is the excluded area for placement of the center of
a rod given that it may not intersect a rod already on the surface. 
This area, averaged over
the isotropic orientational distribution is
\cite{Balberg84b}:
\begin{equation}
A_{exc}= 2 (1+ {4 \over \pi^2}) w \ell + {2 \over \pi} (\ell^2 + w^2).
\label{eq:Perc}
\end{equation}
Note that, since the excluded area 
$A_{exc}$,
is not simply related to the
rod area $\ell \cdot w$, 
the percolation threshold
cannot be expressed as a function of the coverage 
$\eta_2$ alone. 
This distinguishes the rods from isotropic objects such as
disks for which the excluded area is related to the disk
area by a constant factor
\cite{Balberg84b}. 
This difference is especially important in the case of large aspect
ratio of the rods.
In particular, for $w=0$, the coverage vanishes
but the excluded area does not.

The properties of the infinite percolation cluster at the transition have
been studied extensively for lattice systems
as well as
for continuous systems such the the ones considered here
\cite{Stauffer85,Isichenko92}.
In particular, its fractal nature was established. The fractal 
dimension, which can be expressed in terms of other critical 
exponents was found to be universal
\cite{Isichenko92,Kapitulnik84}.
Properties such as 
correlation functions and cluster size distributions  
were also examined
\cite{Torquato88,Lee88,Lee89}.
In addition to percolation systems, spatial fractal structures have 
been observed in a great variety of systems in physics
\cite{mandelbrot82,stanley86,Feder90,Takayasu90,Bunde94,Barabasi95} 
and chemistry 
\cite{avnir92}.

In this paper we perform multiple resolution analysis of
systems composed of randomly adsorbed rods below the
percolation threshold
$\rho_c$. 
In many empirical systems of this type, 
where adsorption and aggregation
processes take place, fractal structures
appear
over a range of length scales
for densities  
below 
$\rho_c$
\cite{me:D}.
Moreover, a broad spectrum of fractal dimensions 
has been observed experimentally in such systems.
To gain a better understanding 
of empirical fractal structures composed of rod-like
particles 
we study these structures using multi-scale analysis, 
based on the 
box-counting (BC)
function. 
In this analysis we examine the fractal properties of the 
entire system rather than
a particular cluster.
This procedure is the one most commonly used in experimental studies 
of adsorption and aggregation phenomena.
In our analysis, we 
concentrate on a range of rod densities below the 
percolation threshold,
where the area fraction covered by the rods is small. 
We identify two ranges of length scales, in which the observed 
dimension of the system is different from the space dimension.
For length scales larger than the typical rod length,
the results resemble
those obtained for randomly distributed spheres
\cite{me:random-model}.
An apparent fractal behavior is observed within a range of length
scales between the rod length and the average distance between
adjacent rods, which can span up to two decades.
However, as was found before in percolation systems, 
highly anisotropic particles exhibit new and important features
which do not appear in the study of isotropic particles
such as spheres.
In particular, we observe that, unlike the case of spheres,
the upper cutoff (and thus the range over which apparent fractal
behavior appears) cannot be expressed in terms of the
effective coverage alone, but also depends
on the excluded area.
Moreover, we find that the apparent fractal dimension (FD)
depends on the orientational distribution and decreases as it
becomes more anisotropic.
For length scales below the rod length,
the BC
function is dominated by the structure
of the single rod 
whether simple or itself fractal. 
The class of systems which exhibits these two ranges 
includes the broad class of systems in which 
fractal-like clusters are adsorbed on a surface.
To examine the fractal properties of such systems we
study a simple one dimensional model in which the
randomly distributed objects are Cantor sets. 
The paper is organized as follows. The model of randomly distributed
rods is introduced in Section II and the box-counting function is
examined. Rods of zero width are studied in Section III,
randomly distributed Cantor sets in Section IV 
and a summary is presented in Section V.

\section{The General Model of Randomly Adsorbed Rods}
\label{rectangles}

In this Section we study a rather general 
model in which $M$ rectangular rods 
are randomly adsorbed on a square surface of area $L^2$.  
They are adsorbed with no correlations and are allowed 
to overlap with each other. 
The lengths of the rods 
$0<\ell_i<\ell_{max}$, 
$i=1,\dots,M$ are independently picked from the 
distribution
$P_{\ell}(\ell)$ and the widths
$0<w_i<w_{max}$, $i=1,\dots,M$ are picked from the   
distribution
$P_w(w)$. 
The orientations 
$0<\theta_i<\pi$ 
are picked from the distribution 
$P_{\theta}(\theta)$.

In order to
examine the apparent fractal nature of the 
resulting random structures we apply the 
BC procedure.  
In this procedure one partitions the surface into a
grid of mesh size $r \cdot L$ ($0<r<1$)  
and counts the number of boxes, $N(r)$,
intersected by at least one rod, as a function of $r$.  An analytical
solution for the BC function $N(r)$ will now be obtained, and its slope
examined in the different domains.

\noindent 
Imagine the surface of area $L^2$ to be initially empty and
arbitrarily choose one of the grid-boxes
of size
$r \cdot L$. 
Denote the excluded area for
placement of the center of rod $i$, given that the rod may not
intersect the chosen box, 
by $S_r(\ell_i,w_i,\theta_i)$. 
This area is shown
shaded in Fig. \ref{fig:excluded-area}(a). 
Simple geometrical arguments yield:

\begin{equation}
S_r(\ell_i,w_i,\theta_i) = (r \cdot L)^2 + \ell_i\cdot w_i + r \cdot L
(\cos\theta_i + \sin\theta_i) (\ell_i + w_i) .
\label{eq:S}
\end{equation}

\noindent 
The two-fold rotation symmetry of the rods indicates that the
angles $\theta_i$ are distributed in the range
$0 \leq \theta_i \leq \pi$. 
However, the four-fold symmetry
of the box makes  
the problem
invariant under the transformation 
$\theta \rightarrow \theta+ \pi/2$.
Therefore, we can simplify the calculations by considering
only the range
$0 \leq \theta_i \leq \pi/2$ 
with no effect on the results.
The angle and side-lengths distributions are thus 
normalized so that:
\begin{eqnarray}
\int_0^{\pi/2} d\theta \:P_{\theta}(\theta) &=& 1 \nonumber \\
\int_0^{\ell_{max}} d \ell \: P_{\ell}(\ell) 
&=& \int_0^{w_{max}} dw \: P_w(w) = 1 .
\label{eq:P}
\end{eqnarray}
We denote quantities averaged with respect to these distributions by
angular brackets: $\langle \cdots \rangle$. 

Following Ref.
\cite{torquato83} 
we consider the probability $q_1$ for
random placement of the first rod without intersecting the box.
This probability,  
which is proportional to the free area,
is given by: 
$q_1 = [L^2-S_r(\ell_1,w_1,\theta_1)]/L^2$. 
The next
rod is placed with new random angle and side-lengths, so that the
probability for two successful placements is:
$q_2=[1-S_r(\ell_1,w_1,\theta_1)/L^2][1-S_r(\ell_2,w_2,\theta_2)/L^2]$.
Clearly, 
$q_M=\prod_{i=1}^M\left[1-S_r(\ell_i,w_i,\theta_i)/L^2\right]$ is
the probability of placing $M$ rods without intersection with the chosen
box. Thus, the probability of at 
least one intersection after $M$ placements
is $p_M=1-q_M$. Since the total number of boxes is $1/r^2$, for a given
realization of angles and side-lengths 
the expected number of intersected
boxes is $p_M/r^2$, or:

\begin{equation}
N_{\{\theta_i,\ell_i,w_i\}}(r) =
{1 \over r^2} \left[ 1 - \prod_{i=1}^M \left( 1-{{
S_r(\ell_i,w_i,\theta_i)} \over L^2} \right) \right] .
\label{eq:N-theta_i}
\end{equation}

\noindent This expression still has to be averaged 
over the side-lengths and
angle ensembles:

\begin{eqnarray}
N(r)=\langle N_{\{\theta_i,\ell_i,w_i\}}(r) \rangle \ \ \ \ \ 
\ \ \ \ \ \ \ \ \ \ \ \ \ \ \ \ \ \ \ \ \ \ \ \ \ \ \ \ \ \ 
\ \ \ \ \ \ \ \ \ \ \ \ \ \ \ \ \ \ \ \ \ \ \ \ \ \ \ \ \ \ 
\ \ \ \ \ \ \ \nonumber \\
={1 \over r^2}\left\{ 1-\left[ \int_0^{\pi/2}d\theta \: 
P(\theta) \int_0^{\ell_{max}} d \ell 
\int_0^{\ell_{max}} dw \: P_{\ell}(\ell) P_w(w)
\left(1-{{S_r(\ell,w,\theta)} \over L^2}\right) \right]^M \right\} .
\label{eq:N-intermediate}
\end{eqnarray}

\noindent 
where the last equality follows since the $\theta_i$ are
independent, identically distributed random variables, 
and so are $\ell_i$ and $w_i$. 
Finally, 
using Eq.
(\ref{eq:S})
this can be expressed in terms of averages:

\begin{equation}
N(r) = {1 \over r^2} \left\{ 1-\left[ 1 - r^2 - 
{{\langle \ell \rangle \langle
w \rangle} \over L^2} - r (\langle \cos\theta + \sin\theta \rangle)
{{\langle \ell \rangle + \langle w \rangle} \over L} \right]^M \right\}.
\label{eq:N-rectangles}
\end{equation}

\noindent 
This BC function contains the information required for
the multi-scale analysis of the system.
In order to proceed with the
scaling analysis, we turn next to the identification of the cutoffs,
which separate between length scales in which $N(r)$ is dominated
by the distribution vs. the internal structure of the rods. 
These cutoffs
are determined by the effective dimensions of the rods,
when projected onto the grid of boxes.
Now, for a given rod $i$, which is oriented at an angle $\theta_i$,
the average 
between the projections along the 
$x$ and $y$ axes (parallel to the box sides) 
is $\ell_i (\cos\theta_i + \sin\theta_i)/2$
for the length
and $w_i (\cos\theta_i + \sin\theta_i)/2$
for the width
[Fig. \ref{fig:excluded-area}(b)]. 
To obtain the effective cutoff, this should
be averaged over all angles and side-lengths. 
We thus expect to find cutoffs
at:

\begin{eqnarray}
r_0^{\ell} &=& \langle \cos\theta + \sin\theta \rangle 
{{\langle \ell \rangle} \over 2L} \nonumber \\
r_0^w &=& \langle \cos\theta + \sin\theta \rangle 
{{\langle w \rangle} \over 2L} 
\label{eq:cutoffs}
\end{eqnarray}

\noindent
Unlike the lower cutoffs which are determined by
the rod dimensions,  
the upper cutoff ($r_1$) is given by
the average distance between adjacent rod sides 
\cite{me:random-model}: 
\begin{equation}
r_1 = {1 \over \sqrt{M}} - \langle \cos\theta + \sin\theta
\rangle {{\langle \ell + w \rangle} \over 2L}
\label{eq:upper_cutoff}
\end{equation}

To gain insight into the behavior of Eq. 
(\ref{eq:N-rectangles})
we display it in 
Fig. 
\ref{fig:rectangles}
for the case of a random distribution of identical 
rods of length $\ell=d_1$ and width $w=d_2$ ($d_1 \gg d_2$) with
isotropic orientations. 
We chose the case of narrow rods, since virtually all of 
the experimentally relevant cases mentioned
above, belong to this category.  
In the next section we return to analyze the
non-isotropic and polydispersed cases. 
The non-trivial apparent fractal region, 
is clearly seen in Fig. 2, between the
predicted cutoffs of 
$\log_{10} (r_0^{\ell}) = -4.2$
and 
$\log_{10} (r_1) = -1.5$.  
It is the range which is typical to
all (fractal) resolution analyses, 
and which we claim, reveals apparent fractality. 
The apparent FD for this region (see derivation below) is 
$D=0.06$. 
In 
Fig.  
\ref{fig:rectangles},
all four different approximately straight line regions are
clearly identified, corresponding to the presence of the three
cutoffs. 
The range $r<r_0^w$, is usually
uninteresting from the experimental point of 
view because it probes the
structure of the building block itself . 
It has a -slope (dimension) of 2,
namely the dimension of the underlying plane.
When $r_0^w<r<r_0^{\ell}$, the
resolution of observation is coarser, 
and the rods appear as
one-dimensional objects, 
reflected in the -slope (dimension) of approximately 1. 
The width of this range depends on the aspect ratio of the rods
and is approximately 
$\log_{10} (\langle \ell \rangle / \langle w \rangle)$ 
in decades.

Having the cut-offs at hand, we now perform a 
scaling analysis in the region between them. 
This will be done by applying
the standard fractal procedure
\begin{equation}
\log_{10} N(r) \sim - D \cdot \log_{10}(r) 
\label{eq:N=rtoD}
\end{equation}

\noindent 
where $D$ is the apparent fractal dimension (FD).
As stated above, the range which has been the 
focus of attention is the one in between the
cutoffs $r_0^{\ell}< r <r_1$. 
The apparent FD, given by the slope in this range, 
is a non-universal dimension, the
magnitude of which depends on the coverage.  
In order to obtain expressions for the coverage dependence, 
it is convenient to define
effective one dimensional (1D) and two dimensional (2D)
coverages. The effective 2D coverage is 

\begin{eqnarray}
\eta_2 = M {{\langle \ell \rangle \langle w \rangle} \over L^2}.
\label{eq:eta2}
\end{eqnarray}

\noindent
while the  effective 1D coverage is 

\begin{eqnarray}
\eta_1 = \eta_1^{\ell} + \eta_1^w
\label{eq:eta1}
\end{eqnarray}

\noindent
where
$\eta_1^{\ell} = \sqrt{M} r_0^{\ell}$
and
$\eta_1^w = \sqrt{M} r_0^w$
(note that as the rod widths approach zero 
$\eta_1 \rightarrow \eta_1^{\ell}$).
The significance of the effective 1D coverage is that
it provides a unified measure of coverage which is independent of 
the space dimension.

The width of the apparent fractal range, 
can be estimated by
$\Delta_e= \log_{10} (r_1) - \log_{10} (r_0^{\ell})$.
In the case of large aspect ratio
($ \langle w \rangle \ll \langle \ell \rangle $):

\begin{equation}
\Delta_e = 
\lim_{\langle w \rangle/\langle \ell \rangle  \rightarrow 0} 
(\log r_1 - \log r_0^{\ell}) \approx \log (1/\eta_1 -1). 
\end{equation}

Since in the limit of large aspect ratio
$\eta_1 \propto (\rho \cdot A_{exc})^{1/2}$
up to a factor of order 1, 
it is found, 
from the criterion of
Ref. 
\cite{Balberg84b}
($\rho_c \cong 1/A_{exc}$)
that the width 
$\Delta_e$
is nearly zero in the vicinity of the percolation
threshold, and increases as the density decreases below
this threshold.
This estimate for 
$\Delta_e$
may seem to suggest that one can increase the scaling
range at will by decreasing the coverage. 
However, in addition to the width of the range between the
cutoffs, the quality of the linear fit within this 
range measured by the coefficient of 
determination $R^2$ should be considered.
One can limit the range of linearity by imposing a 
lower bound on $R^2$.
Obviously, the range decreases as $R^2$ increases.
We thus conclude that the two cutoffs 
limit the width of the linear range for high coverage
while the $R^2$ criterion limits it for low coverage
[see
Ref. \cite{me:random-model} and compare 
Figs. \ref{fig:BC-needles}(a) and \ref{fig:BC-needles}(b)]. 
The net result is that
the range of scaling is in fact restricted to 1-2 decades.

The apparent dimension is found by calculating the logarithmic
derivative of $N(r)$ at the estimated middle point 
of the linear range. This
point is situated at:
$r_m = \sqrt{r_1 \cdot r_0^{\ell}}$.
Assuming $\langle w \rangle < \langle \ell
\rangle$ this yields for the apparent FD, in the $M 
\rightarrow \infty$ limit and at constant coverage, the general equation:

\begin{equation}
D = 
{{d \log[N(r)]} \over {d \log(r)}}\left|_{r_m}\right. = 
2 \left[ 1 - {{(1-\eta_1)\eta_1^{\ell} + 
\eta_1\sqrt{(1-\eta_1)\eta_1^{\ell}}} \over
{\exp[\eta_1^{\ell}(1-\eta_1)+
2\eta_1\sqrt{(1-\eta_1)\eta_1^{\ell}}+\eta_2]-1}} \right]
\label{eq:D-rects}
\end{equation}

\noindent 
This general formula for the FD will be used in the
analysis to follow.
The effects of changing the coverage on the apparent dimension, 
are shown in
the inset of Fig. 
\ref{fig:rectangles}
for $d_1/d_2 = 10^3$: the FD increases
monotonically with coverage, but does not reach 2 for the 
relatively narrow
rods considered.
 
The dimension of the rods themselves can also be obtained 
from the box counting function. In order to do so one needs
to identify the appropriate length scale and measure the 
slope of the box counting function
$N(r)$ 
on the $\log-\log$ plot.
This slope should be obtained 
for a length which is the geometrical average of the effective
cutoffs associated with the average rod length 
$\langle \ell \rangle$ and width 
$\langle w \rangle$
according to
$r=\sqrt{r_0^{\ell} \cdot r_0^w}$.
In the zero-width rod limit $d_2/d_1 \rightarrow 0$
this dimension approaches 1. 
As we are focusing especially on the experimentally 
relevant case of narrow
rods, and as it is evident from Eqs.
(\ref{eq:eta2}), (\ref{eq:eta1})
that both the FD and the
range $\Delta_e$ are only marginally dependent on the rod width,  
we consider   
in the next  cases, for simplicity,  rods with
zero width.

\section{Applications to Rods of Zero Width}
\label{needles}

In the present Section we consider for simplicity the case of rods with
zero width. Some comparisons with simulations will be presented, 
as well as
applications of the general theory to specific distributions of interest.

\subsection{Isotropically Oriented Identical Rods}
\label{simul}
As a first case for the zero-width rods, let us return 
to the previous example
of identical rods with isotropic orientations
and impose zero width ($w=0$).
For the isotropic 
$[P_{\theta}(\theta)=2/\pi]$, 
monodispersed $[P_{\ell}(\ell)=\delta(\ell-d)]$
case, the BC function and FD for rods of length $d$ and zero 
width follow
directly from Eqs.
(\ref{eq:N-rectangles}),
(\ref{eq:D-rects}) respectively:

\begin{equation}
N(r) = {1 \over r^2} \left\{ 1-\left[ 1- r^2
- r {4 \over \pi} {d \over L} \right]^M \right\} ,
\label{eq:BC-needles}
\end{equation}

\begin{equation}
D = 2 \left[ 1 - {{\eta_1^{\ell}
(1 + \sqrt{(1-\eta_1^{\ell})\eta_1^{\ell}} -\eta_1^{\ell})} \over
{\exp[\eta_1^{\ell}(1+2\sqrt{(1-\eta_1^{\ell})
\eta_1^{\ell}}-\eta_1^{\ell})]-1}} \right] .
\label{eq:D-needles}
\end{equation}

\noindent 
In 
Fig. \ref{fig:BC-needles} 
we present this analytical result for the
rod-BC function along with numerical simulations, 
for two coverages.  The
agreement between theory and simulations is 
excellent over the entire range of
box-sizes.  The ranges of apparent fractality are 
brought in the insets; one
would expect the wavy nature of the line at the lower coverage, to be
smeared out by noise in typical experimental situations.  
Comparing to the BC function for
finite-width rods, 
which appears in
Fig. \ref{fig:rectangles}, 
one observes that, as
expected, there are now only two cutoffs and correspondingly three
(approximately) linear regions of slope $\sim$ 1, $D<1$ and 2.

\subsection{Anisotropically Oriented Identical Rods}
\label{anisotropy}

Anisotropically oriented elongated particles appear in a wide variety of 
systems, notably in liquid crystals. 
In order to investigate the effect of 
anisotropy on the apparent fractal properties 
we consider here the following 
angular distribution, normalized for $0 \leq \theta \leq \pi/2$:

\begin{equation}
P(\theta) = 
{{2 \Gamma(1+n)} \over {\sqrt{\pi} \Gamma(1/2+n)}} (\cos \theta)^{2n}.
\label{eq:P(theta)}
\end{equation}

\noindent In the limit $n \rightarrow 0$ this corresponds to a 
uniformly random distribution, whereas for $n \rightarrow \infty$, to
perfectly aligned rods. The rods are assumed to be of
equal size $d$. The BC function can be found by calculating the angular
averages of Eq.
(\ref{eq:N-rectangles}). 
Using the identities ($n > 0$):

\begin{eqnarray}
\int_0^{\pi/2} d\theta \: \cos^{2n+1}\theta = 
{{\sqrt{\pi} \Gamma(1+n)} \over {2 \Gamma(3/2+n)}} \\
\int_0^{\pi/2} d\theta \: \cos^{2n}\theta \sin\theta = {1 \over 1+2n} ,
\label{eq:ident}
\end{eqnarray}

\noindent we find:

\begin{equation}
\gamma_n \equiv (\langle \cos\theta + \sin\theta \rangle) = {\Gamma(1+n)
\over \Gamma(1/2+n)} \left( {\Gamma(1+n) \over 
\Gamma(3/2+n)} + {2 \over {\sqrt{\pi}(1+2n)}} \right).
\label{eq:gamma_n}
\end{equation}

\noindent This yields for the BC function:

\begin{equation}
N(r) = {1 \over r^2} \left\{ 1 - \left[ 1 - r^2 - r {d \over L} \gamma_n
\right]^M \right\}.
\label{eq:N-anisotropic}
\end{equation}

\noindent In the $n \rightarrow 0$ limit one retrieves the result for 
uniformly randomly oriented rods, Eq.
(\ref{eq:BC-needles}), whereas for 
$n \rightarrow \infty$ one finds 
$N(r) = \{ 1-[ 1 -r^2 - r \, d/L ]^M \}/r^2$. 
For the cutoffs we have from Eq.
(\ref{eq:cutoffs}):

\begin{eqnarray}
r_0 &=& \gamma_n {d \over 2L}
\nonumber \\
r_1 &=& {1 \over \sqrt{M}} - \gamma_n {d \over 2L} ,
\label{eq:cutoffs'}
\end{eqnarray}

\noindent and for the coverage:

\begin{equation}
\eta_1^{\ell} = \sqrt{M}  \gamma_n {d \over 2L}
\label{eq:eta12'}
\end{equation}

\noindent The FD for the present case is found by substituting 
$\eta_1^{\ell}$ from Eq.
(\ref{eq:eta12'}) in the general expression
Eq.
(\ref{eq:D-rects})
and taking
$\eta_2=0$.
The effect on $D$ of changing $n$ at constant number
and size of rods is shown in 
Fig. \ref{fig:D-anisotropic}: 
the apparent FD
decreases as the rods become more parallel, i.e., as they ``cover space''
less effectively.
For isotropic objects we have shown
that the apparent FD depends essentially only on the coverage 
\cite{me:random-model}.
As seen here, for anisotropic objects, the FD depends on an 
additional parameter (the degree of anisotropy). 
This feature may be relevant to many experimental systems which
exhibit anisotropic distributions of rod-like particles, such as
liquid crystals and paper fibers.

\subsection{Polydispersed Rods}
\label{poly}

We will now explore the effects of polydispersivity in the rods size. In 
particular we will consider a power-law distribution 
of the rod lengths.

\subsubsection{Model}
\label{poly-model}
In Ref.\cite{me:random-model} a variety of narrow size distributions were
examined. 
It was observed that such polydispersivity does not alter the basic
observation of an apparent fractal regime between cutoffs. The apparent FD
of the corresponding monodispersed distribution was only slightly modified.
An important distribution function found in numerous experimental 
cases is the
power-law distribution of sizes 
\cite{power_laws}.  
To obtain such a distribution we
choose rod lengths from an iteratively constructed Cantor set, containing
$2^n$ segments of length $3^{-n} L$ in the $n^{\rm th}$ iteration. Assuming
rod lengths are chosen uniformly from among these segments, the probability
of choosing a segment of length $3^{-n} L$ is 
$P(\ell/L = 3^{-n}) = 2^n/Z$, where
$Z = \sum_{k=1}^{k_m} 2^k$, $k_m$ being the maximal iteration number in the
construction of the Cantor set. 
The average rod length $\ell$ is then given
by:

\begin{equation}
\langle \ell \rangle = \sum_{k=1}^{k_m} \ell\,P(\ell/L=3^{-k}) = {1 \over Z}
\sum_{k=1}^{k_m} 2^k 3^{-k} L = {{1 - (2/3)^{k_m}} \over {2^{k_m} - 1}} L .
\label{eq:<l>}
\end{equation}

\noindent Expressed in terms of lengths one finds for the distribution:

\begin{equation}
P(\ell) = {1 \over Z} 2^{-\log(\ell/L)/\log(3)} = 
{1 \over Z} (\ell/L)^{-D_c} \:\: ;
\:\:\:\: D_c = {\log(2) \over \log(3)} ,
\label{eq:P(l)}
\end{equation}

\noindent $D_c$ being the FD of the Cantor set. Thus the length
distribution indeed satisfies a power-law \cite{D-comment4}. 
Note that $P(\ell)$
is a {\em discrete} distribution with allowed lengths of 
$\ell = 3^{-n} L$, and
that $\langle \ell \rangle$, 
for example, would be different if $\ell$ could assume
any value between $3^{-k_m}L$ and $L/3$.

\subsubsection{Scaling Analysis for Power-Law Dispersed Rods}
\label{scaling-needles}

We assume that the rods are uniformly randomly oriented, so that $\langle
\cos\theta + \sin\theta \rangle = 4/\pi$,
$0 \le \theta < \pi/2$. 
We thus have from
Eq.
(\ref{eq:N-rectangles}):

\begin{equation}
N(r) = {1 \over r^2} \left\{ 1-\left[ 1 - r^2 - 
r {4 \over \pi} {{\langle \ell
\rangle} \over L} \right]^M \right\} 
\label{eq:N-needles}
\end{equation}

\noindent with $\langle \ell \rangle$ 
given by Eq.
(\ref{eq:<l>}). 
As can be seen in 
Fig. \ref{fig:BC-Cantor-needles}, 
for as few as
$M=62$ rods, Eq.
(\ref{eq:N-needles}) is in good agreement with the
simulation results for rods with a power-law distribution of lengths.
Thus, we find that the BC function can be derived analytically for
a power-law distribution of particle 
dimensions, and depends essentially only
on the first moment of this distribution.
The apparent FD can also be found as before, as the slope between cutoffs
determined by the average rod length. From Eq.
(\ref{eq:cutoffs}), one gets
in the zero-width rod limit:

\begin{eqnarray}
r_0 &=& {2 \over \pi} {{\langle \ell \rangle} \over L} \nonumber \\
r_1 &=& {1 \over \sqrt{M}} - {2 \over \pi} 
{{\langle \ell \rangle} \over L} .
\label{eq:cutoffs-needles}
\end{eqnarray}

\noindent The effective 1D coverage is now given by:

\begin{equation}
\eta_1^{\ell} = \sqrt{M} {2 \over \pi} {{\langle \ell \rangle}  \over L} .
\label{eq:eta1-needles}
\end{equation}

\noindent The FD is found from Eq. 
(\ref{eq:D-needles}), with the present
$\eta_1^{\ell}$. 
The
prediction of this formula is compared in 
Fig. 
\ref{fig:BC-Cantor-needles} with
a linear regression in the range set exactly by the above cutoffs, and is in
good agreement. Thus the FD is still determined essentially only by the 1D
coverage. This is a non-trivial result, since a power-law distribution is
poorly described by its mean, yet this is essentially the only
distribution-related quantity needed to express the BC function. It can be
understood intuitively as follows: since the BC function counts the total
number of occupied boxes, it is approximately
proportional to the total length of all
rods. This quantity is well described by the number of rods times the
average rod-length, namely the 1D coverage. In the next Section we consider
the effect of endowing the rods with an internal (fractal) structure.
\section{Randomly Adsorbed Cantor-Rods}
\label{Cantor}

So far we have dealt with random distributions of objects which are not
themselves fractals. This was reflected in the BC function at resolutions
below the lower cutoff, by an integer slope. In this Section we consider a
model of randomly deposited rods which are all Cantor sets of FD=$D_c$. 
A new
feature expected in this case, is that for perfect (i.e., not truncated)
Cantor sets, a {\em non-integer}
slope of $-D_c$ should appear below the lower cutoff, in contrast to the
cases considered 
so far. The main motivation for considering a model of Cantor-rods, however,
is that it mimics a large class of experimental systems where a set of
fractal objects is randomly adsorbed on a surface. 
For example, DLA-like clusters
growing simultaneously from several nucleation centers 
\cite{hwang91}
and cluster-cluster aggregation experiments
\cite{meakin88}. 
In such
systems we expect an interplay between the FD of the fractal objects and the
apparent FD induced by randomness. In particular, it might be difficult to
disentangle the respective slopes of the BC function if the objects are
fractal over a small range.

For simplicity we will now consider the one dimensional case where
$M$ Cantor sets (rods) of total length $d$ each, are deposited on a
line of length $L$. 
As in Sec.\ref{needles}, let $n$
denote the number of iterations in the construction of the Cantor sets. Thus
there are $2^{n}$ segments of length $3^{-n}$ in each rod. 
As before, the rods
are fully penetrable to each other. In order to find the BC function for the
resulting set, we focus on a single, arbitrary box of length $r\cdot L$ and
calculate the excluded length for placement of the center of a rod, the
condition being that the rod does not overlap with the box.

Let $\Omega_k^{(n)}$ denote the excluded length for placement of a single
$n^{\rm th}$ iteration Cantor-rod, when the box-length satisfies: $3^{-k-1} d
< r \cdot L < 3^{-k} d$. To find this function, consider first the case $n=0$,
i.e., the case of gapless rods of length $d$. This is nothing but the 1D
version of equi-sized rods considered in Sec.\ref{rectangles}, and the
excluded length is clearly $\Omega_0^{(0)} = r \cdot L + d$. When $n=1$ there
is a gap of length $d/3$ and one must distinguish between the cases $r \cdot L
> d/3$ and $r \cdot L < d/3$. In the former, the resolution of the boxes is
insufficient to notice the presence of the gap, namely, if the box overlaps
with the gap it necessarily touches at least one of the two rod-segments as
well. In this case, therefore, the excluded length is again $\Omega_0^{(1)} =
r \cdot L + d$. However, when $r \cdot L < d/3$ a new situation arises: the
box can fully overlap with the gap. This is equivalent to having two rods of
length $d/3$, each contributing $r \cdot L + d/3$ to the excluded length:
$\Omega_1^{(1)} = 2(r \cdot L + d/3) $. When $n=2$ there are gaps of length
$d/3$ and $d/9$, so that three cases arise: (1) $r \cdot L > d/3$, (2) $d/9 <
r \cdot L < d/3$, (3) $r \cdot L < d/9$. The first two do not differ from
$n=1$ since the the boxes are not small enough to resolve the $d/9$ gaps:
$\Omega_0^{(2)} = r \cdot L + d$, $\Omega_1^{(2)} = 2(r \cdot L + d/3)$. The
third case is equivalent to having four rods of length $d/9$, each
contributing an excluded length of $r
\cdot L + d/9$: $\Omega_2^{(2)} = 4(r \cdot L + d/9)$. The general case
should now be clear:

\begin{eqnarray}
\Omega_k^{(n)} = \left\{ \begin{array}{ll}
        r \cdot L + d                                                       
& \mbox{: $r > d/L$} \\
        2^k (r \cdot L + 3^{-k} d)                                        
& \mbox{: $3^{-k-1} d/L < r < 3^{-k}
d/L, \: 0 \leq k \leq n$} \\
        2^n (r \cdot L + 3^{-n} d)                                        
& \mbox{: $r < 3^{-n-1} d/L .$}
        \end{array}
\right.
\label{eq:Omega}
\end{eqnarray}

\noindent It is convenient to express the index $k$ 
satisfying the constraint
in Eq. (\ref{eq:Omega}) as:

\begin{equation}
k = \lfloor { {\log[d/(r \cdot L)]} \over {\log(3)} } \rfloor
\label{eq:k}
\end{equation}

\noindent (where $\lfloor x \rfloor$ is the largest integer smaller than
$x$). Now suppose $r$ is given, choose an arbitrary box, and place a
Cantor-rod at random on the line. The probability $q_1$ that the rod does not
intersect the box is the relative available length, i.e., $q_1 =
1-\Omega_k^{(n)}/L$. For $M$ independently placed Cantor-rods the probability
that none intersects the chosen box is $q_M = q_1^M$, and the probability of
at least one intersection is: $p_M = 1 - q_M$. When multiplied by the total
number of boxes ($1/r$), this yields the expected number of 
intersected boxes:

\begin{eqnarray}
N(r) = \left\{ \begin{array}{ll}
	\left[ 1- [ 1- ( r + d/L ) ]^M \right]/r
& \mbox{: $r > d/L$} \\ 
	\left[ 1- \left[ 2^k (r + 3^{-k} d/L) \right]^M \right]/r
& \mbox{: $r < d/L, \: 0 \leq k \leq n$} \\ 
	\left[ 1- \left[ 2^n (r + 3^{-n} d/L) \right]^M \right]/r
& \mbox{: $r < d/L, \: k > n .$} 
	\end{array}
\right.
\label{eq:N-Cantor}
\end{eqnarray}

\noindent where $k$ is given by Eq.
(\ref{eq:k}). This expression for the BC
function is plotted in 
Fig.
\ref{fig:N-Cantor} for fifth generation
Cantor-rods. The remarkable feature in comparison with randomly positioned
full rods is that the lower cutoff has shifted to the left, now reflecting the
size of the smallest segment in the Cantor-rods. At smaller resolutions the
boxes again ``see'' 1D objects and the slope of the BC function is $-1$. At a
resolution of $r=d/L$ (rod-size), there is a smooth transition into the regime
of apparent fractality associated with the random distribution of
Cantor-rods. We believe that such behavior is typical of experimental
situations where a random distribution of limited-range fractal objects is
observed. However, unlike the present ``clean'' case, it may be much more
difficult to separate the two regions in actual experimental data.

It should further be remarked that as can easily be checked,
Eq.
(\ref{eq:N-Cantor}) 
yields a logarithmic derivative for $N(r)$ of $-D_c$ in
the limits $n \rightarrow \infty$ and $r \rightarrow 0$. This implies the
interesting result that regardless of coverage, a random ensemble of Cantor
sets has exactly the same FD as a single Cantor set. 
Fig.
\ref{fig:N-Cantor}
shows that in the range between the size of the smallest segment and the size
of the set, the same conclusion holds for truncated Cantor sets. Clearly, our
methods of calculation can be extended to other iteratively constructed
fractal sets, and the conclusions above regarding the FD should therefore
remain valid in such cases. We conjecture that they remain valid also for
non-iteratively constructed fractals.

\section{Summary}

We have performed a multiple resolution analysis using box counting
functions to structures composed of randomly adsorbed rods.
Such structures appear in a large variety of 
adsorption phenomena and in many
physical systems
including liquid crystals, aggregates of linear molecules
and fibers and line fractures. 
The scaling properties of these systems are determined by the
particle size (length and width) distribution, orientation 
distribution, as well as by correlations in the positions and 
orientations. In processes such as paper fiber sedimentation
particles tend to aggregate into dense regions and create
inhomogeneities
\cite{provatas95,provatas96}. 
In other systems, rods cannot overlap
giving rise to a maximal jamming density
\cite{Chick,viot92}. 

We have studied the case where rods can overlap and there are no
positional or orientational correlations between them. We examined
various size distributions and orientation distributions, found an
analytical expression for the box counting function 
and compared the analytical results to numerical simulations.

For rod densities below the percolation threshold,
we identified two interesting ranges of length-scales in which
the box counting analysis gives rise to non trivial scaling 
properties.
In the range of
length scales between the typical rod length and 
the typical distance between adjacent rods,
the box counting
function is determined by the positional 
and orientational distribution of the 
rods in the plane rather than the structure of the single rod.
This gives rise to an apparent fractal behavior 
over a finite range of up to two decades.
Unlike the case of randomly distributed isotropic objects (disks),
the range of length scales over which apparent fractal
behavior is observed depends not only on
effective coverage 
but also on the excluded area
(averaged over the orientational distribution).
Moreover, the apparent fractal dimension
depends on the orientational distribution and decreases as it
becomes more anisotropic.

For length scales smaller than the typical rod length, the box
counting function is determined by the internal structure of the
rod. In case of ordinary one dimensional rods this gives rise
to a dimension of 1. However, this result is more general and
in case that the rods are trimmed into Cantor sets the FD which
is observed in this range is equal to the FD of the single Cantor
set.
The distinction between these two ranges of length scales applies
for a very broad class of systems which exhibit nucleation of 
a finite density of fractal-like clusters. 
The majority of spatial fractals in the physics literature
belong to this class
\cite{me:D}.
A finite density of fractal-like clusters
appears in models such as 
diffusion-limited-cluster-cluster-aggregation 
\cite{meakin88}
and in experimental systems such as molecular beam epitaxy
where diffusion-limited-aggregation-like clusters nucleate
at a finite density
\cite{hwang91}.
In these systems the fractal-like structure of the clusters results 
from complex stochastic dynamics.
The cluster density is determined by parameters such as the 
temperature and deposition rate. The typical distribution is 
not exactly Poissonian due to effective repulsion between clusters. 
However, the distinction,
emphasized here,
between smaller length scales where the 
fractal properties are dominated by the single cluster and larger
length scales where the distribution is dominant still applies. 
We thus predict that experiments involving random distributions
of fractal-like objects, will reveal a cross-over from object-dominated
to distribution-dominated fractal behavior.

\acknowledgments

We would like to thank Prof. R.B. Gerber and D. Thimor for very helpful
discussions. 
This work was supported by a grant from the Wolkswagen Foundation,
administered by the Niedersachsen Science Ministry.
D.A. acknowledges support by the German BMBF and the Minerva Foundation,
Munich.

\begin{figure}
\caption{(a) Excluded area for the placement of the center of a rod 
of length $\ell=d_1$ and width $w=d_2$ inclined at angle $\theta$ relative
to the horizontal axis. 
If the center falls outside the shaded area, the rod will not intersect 
the box of side $r$; 
(b) At the orientation shown, the projections of the sides, 
$d_1$ and $d_2$, on the horizontal axis are 
$d_1 \cos\theta$ and $d_2 \sin\theta$ respectively
(the sum of the two projections is shown by the thick line). 
The projections on the vertical axis are
$d_1 \sin\theta$ 
and 
$d_2 \cos\theta$
respectively. 
Averaging over the two possibilities and $\theta$ yields
Eq. (\ref{eq:cutoffs}).}
\label{fig:excluded-area}
\end{figure}

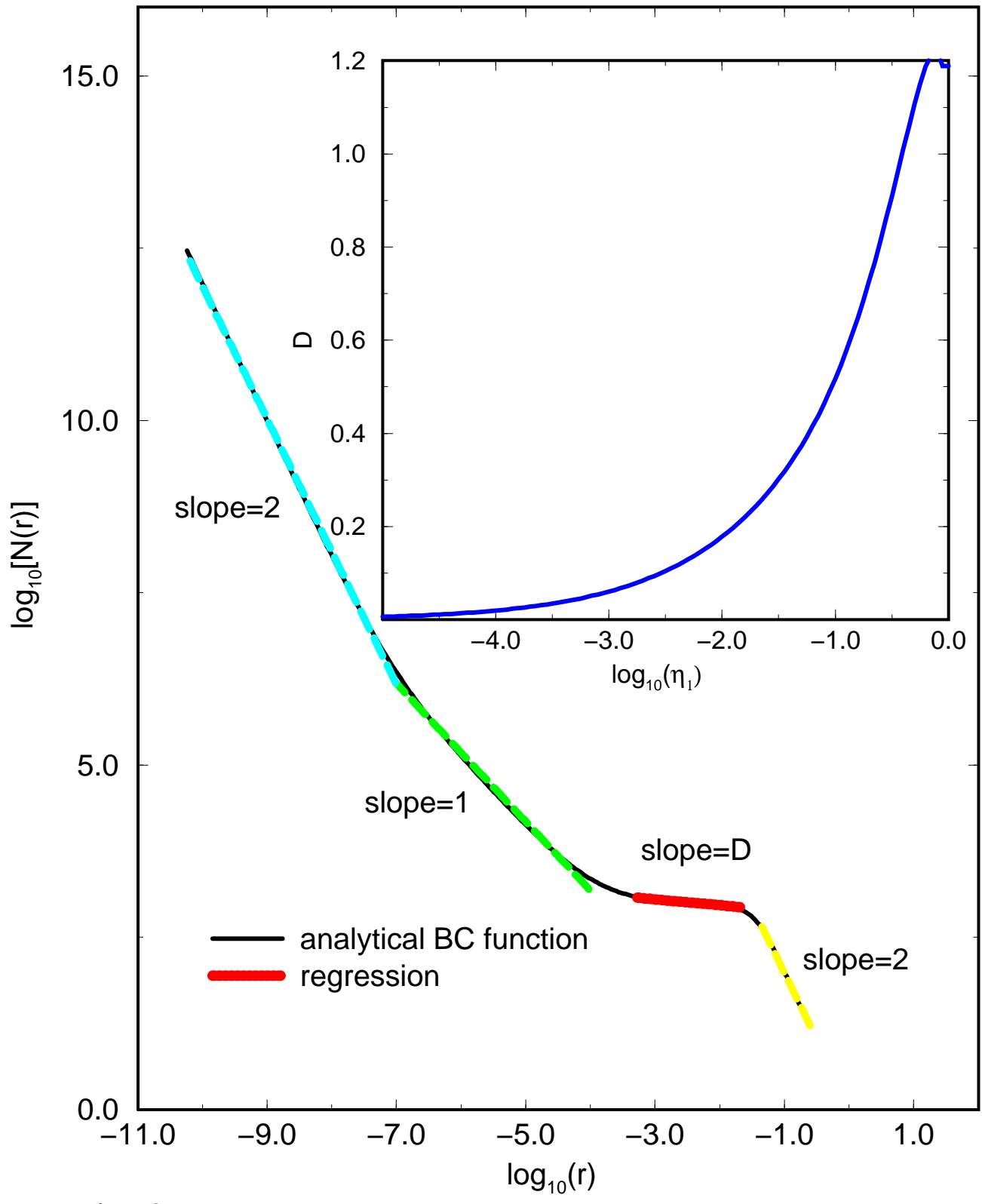
\begin{figure}
\caption{
The BC function for rods 
(Eq.  
\ref{eq:N-rectangles}).
Four approximately linear regimes
indicate the presence of three cutoffs. 
As can be seen, the locations of these
match the prediction of Eq. (\ref{eq:cutoffs}) 
[$\log_{10}(r_0^{\ell})=-4.2$,
$\log_{10}(r_0^w)=-7.2$, 
and $\log_{10}(r_1)=-1.5$].
The inset shows
the apparent fractal dimension $D$, given by  
Eq. (\ref{eq:D-rects}),
(for the range of length scales between 
$r_0^{\ell}$ and $r_1$)
plotted as a function of the effective 1D coverage $\eta_1$. 
$D$ increases monotonically with $\eta_1$. 
} 
\label{fig:rectangles}
\end{figure}

\begin{figure}
\caption{BC functions for zero-width rods. 
(a) $\eta_1^{\ell} = 0.1, d_1 =
0.01$; Upper 
inset: typical configuration of rods; Lower inset: zoom between cutoffs
$\log(r_0)=-2.2$, $\log(r_1)=-1.2$. 
(b) $\eta_1^{\ell} = 0.01, d_1 = 0.001$; Inset:
zoom with agreement on slope between Eq. 
(\protect\ref{eq:D-needles}) and the
linear regression to 2 significant digits.}
\label{fig:BC-needles}
\end{figure}

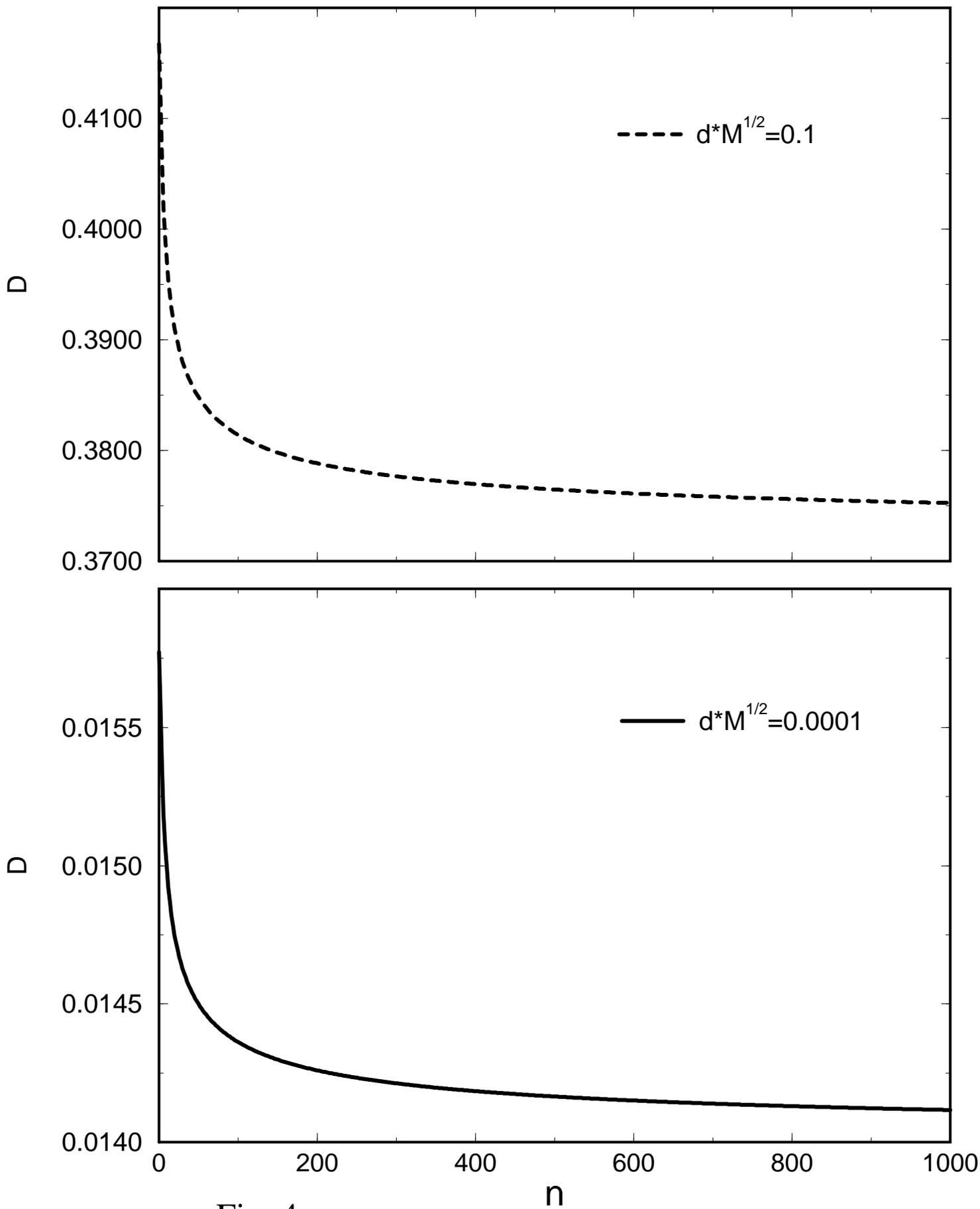
\begin{figure}
\caption{Apparent FD as a function of anisotropy parameter $n$
[Eqs. (\ref{eq:D-needles}), (\ref{eq:eta12'})], 
at two different 1D coverages
($d M^{1/2}$). The FD decreases as the rods tend to be more parallel, and
as the coverage becomes smaller.}
\label{fig:D-anisotropic}
\end{figure}

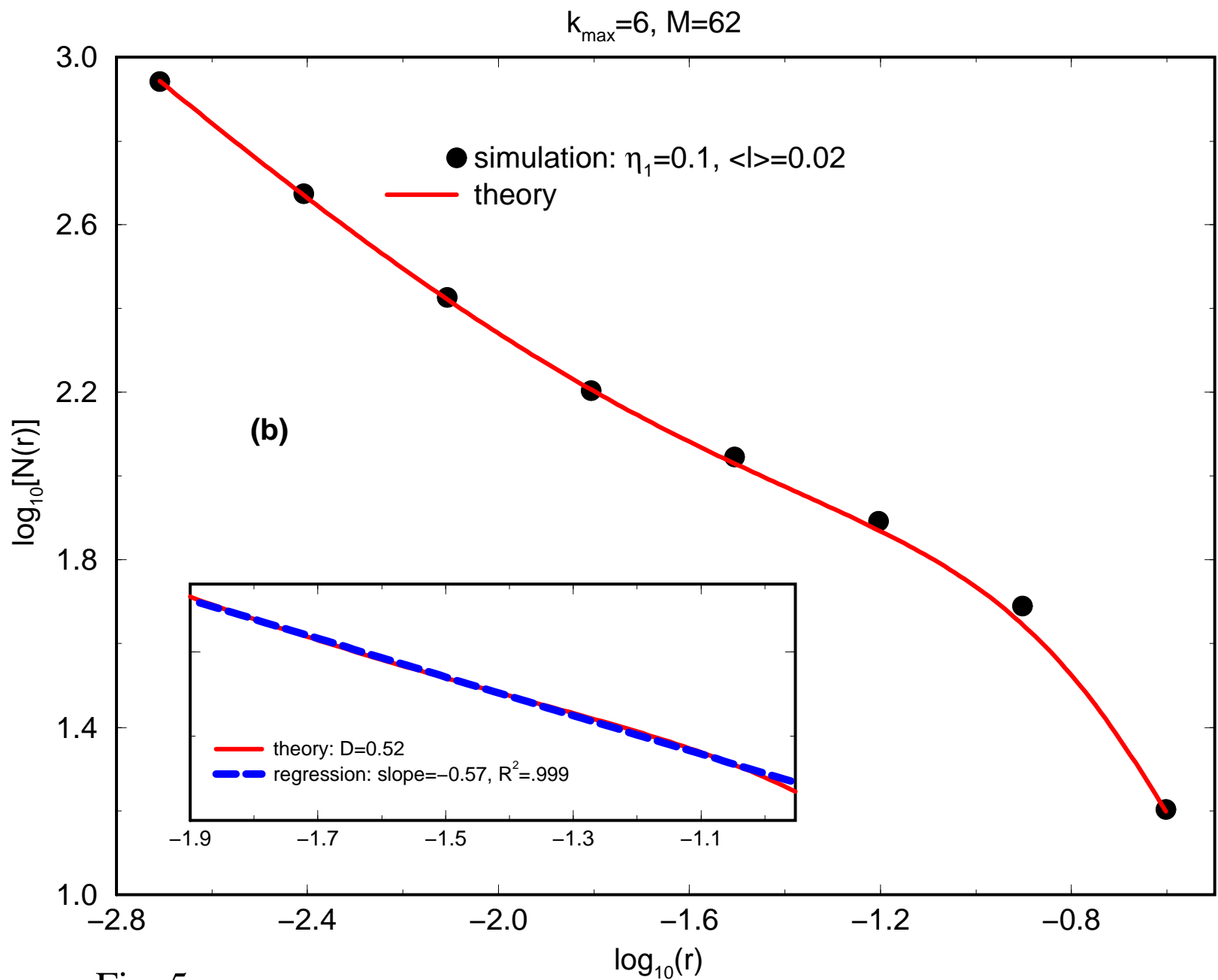
\begin{figure}
\caption{BC functions for isotropically oriented 
rods of zero width and a power law
distribution of lengths. 
The coverage here is
$\eta_1 = 0.1$,
the average rod length is
$\langle l \rangle = 0.02$,
the number of rods is
$M=62$, 
and the number of possible sizes generated 
by Cantor set iterations is
$k_m = 6$. 
The inset shows the
zoom between cutoffs 
$\log_{10}(r_0)=-1.89$ and 
$\log_{10}(r_1)=-0.94$.
}
\label{fig:BC-Cantor-needles}
\end{figure}

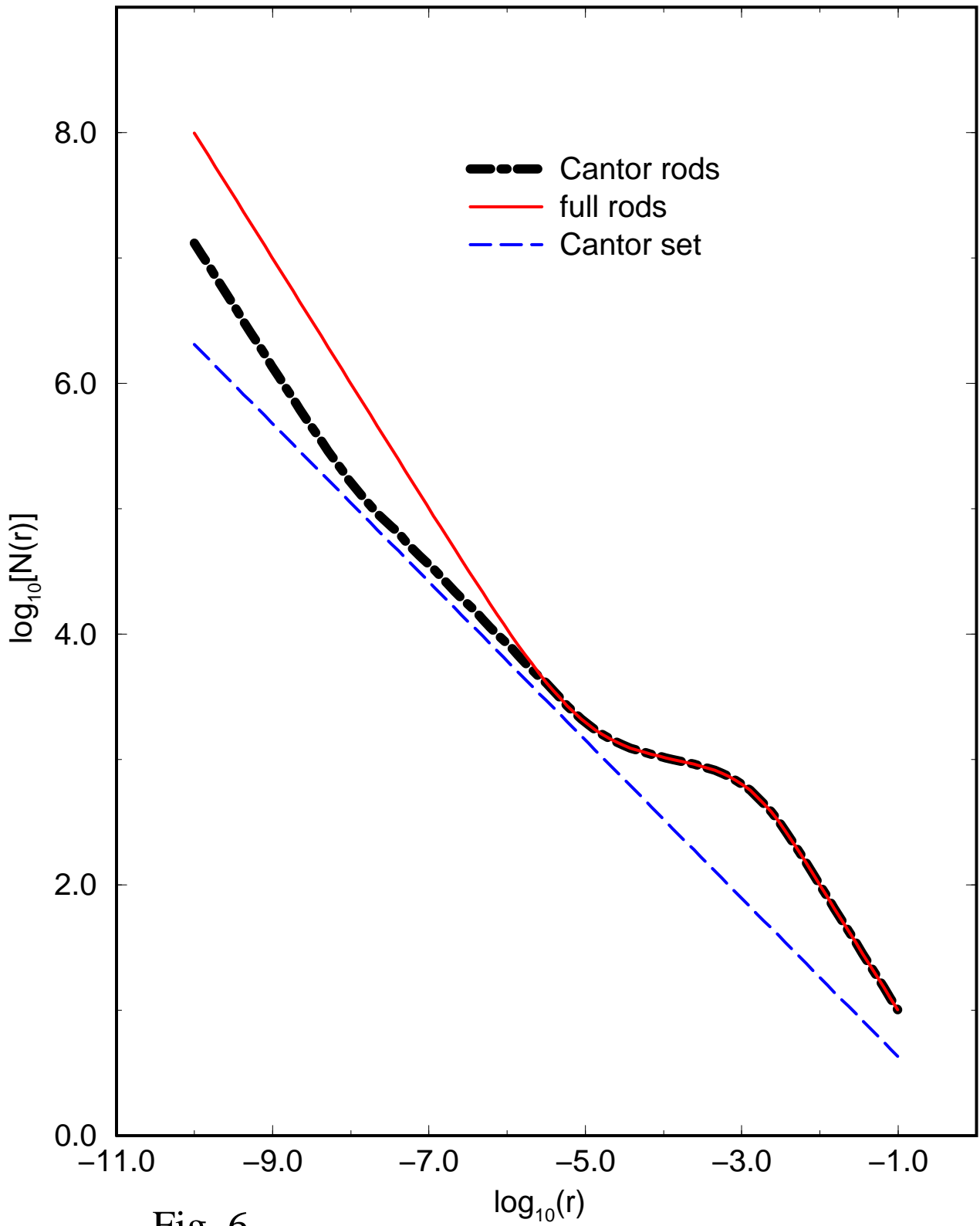
\begin{figure}
\caption{BC function for a random distribution of Cantor-rods
on the interval (thick dashed-dotted line). 
Comparisons to the result for full rods of the same 
total length (solid line) as well as for
a single perfect Cantor set with FD=$D_c$ (dashed line)
are shown. 
The four regions observed
for the Cantor-rods are 
(from left to right)
of slope $-1$, 
slope $-D_c$ (which is the Cantor-set dimension determined by 
the internal structure of the rods), 
apparent FD 
(which is determined by the coverage and the distribution of rods), 
and slope $-1$ again. 
In this example there are $M = 1000$
Cantor rods of size
$d = 10^{-5} \cdot L$ each
and the number of iterations in the construction
of each rod is
$n = 5$.
}
\label{fig:N-Cantor}
\end{figure}


\begin{thebibliography}{10}
\bibitem{D-comment4}
{It is also a hyperbolic distribution \cite{mandelbrot82}, in the sense that
  $P(l/L \geq \lambda) = 2(\lambda^{-D_c} - 1)/Z \sim \lambda^{-D_c}$. This
  follows from $P(l/L \geq 3^{-n}) = \sum_{i=0}^{n-1} P(l = 3^{i-n}) =
  \sum_{i=0}^{n-1} 2^{n-i}/Z = 2(2^n-1)/Z$ and $\lambda = 3^{-n}$.}
\end{thebibliography}
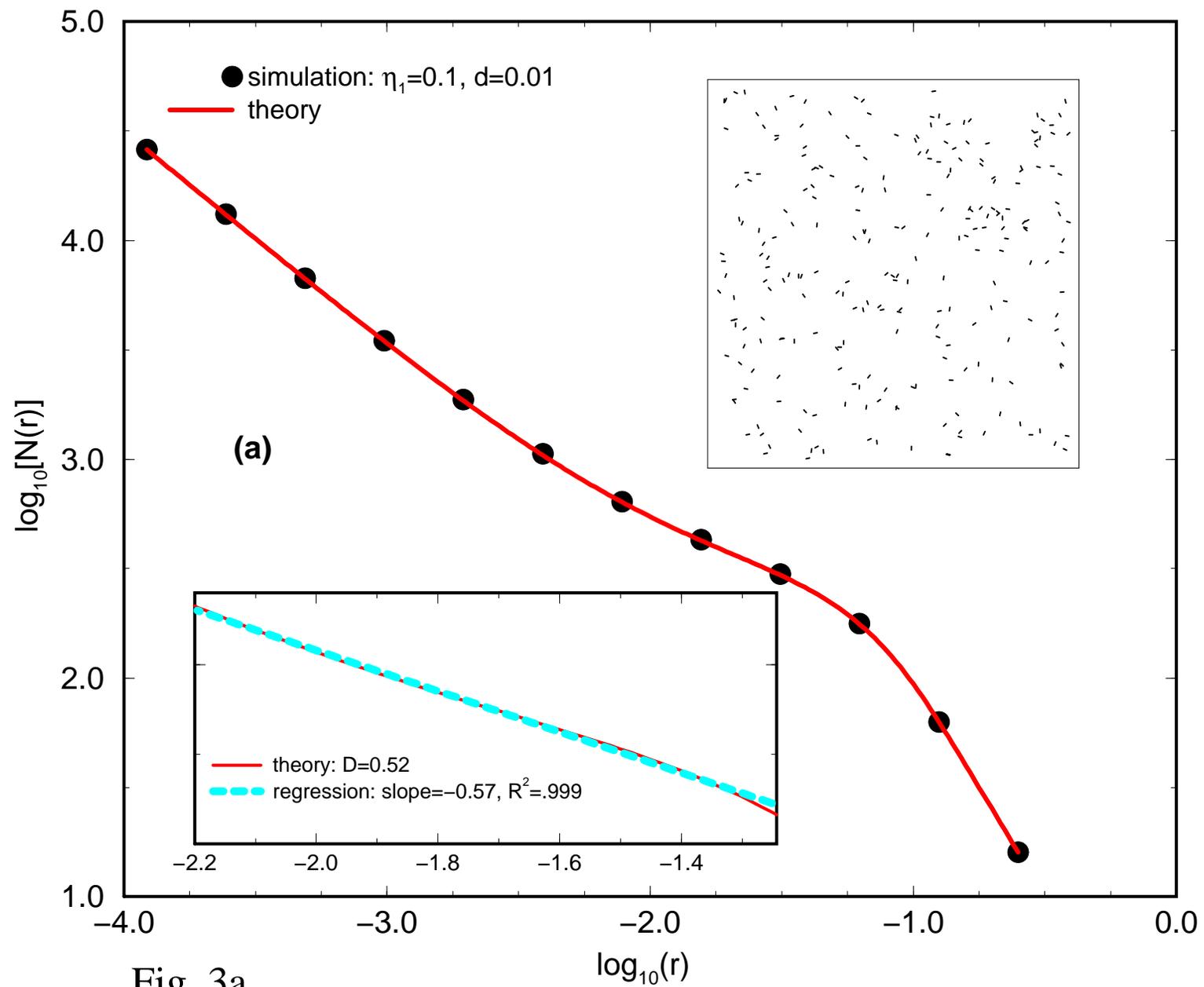
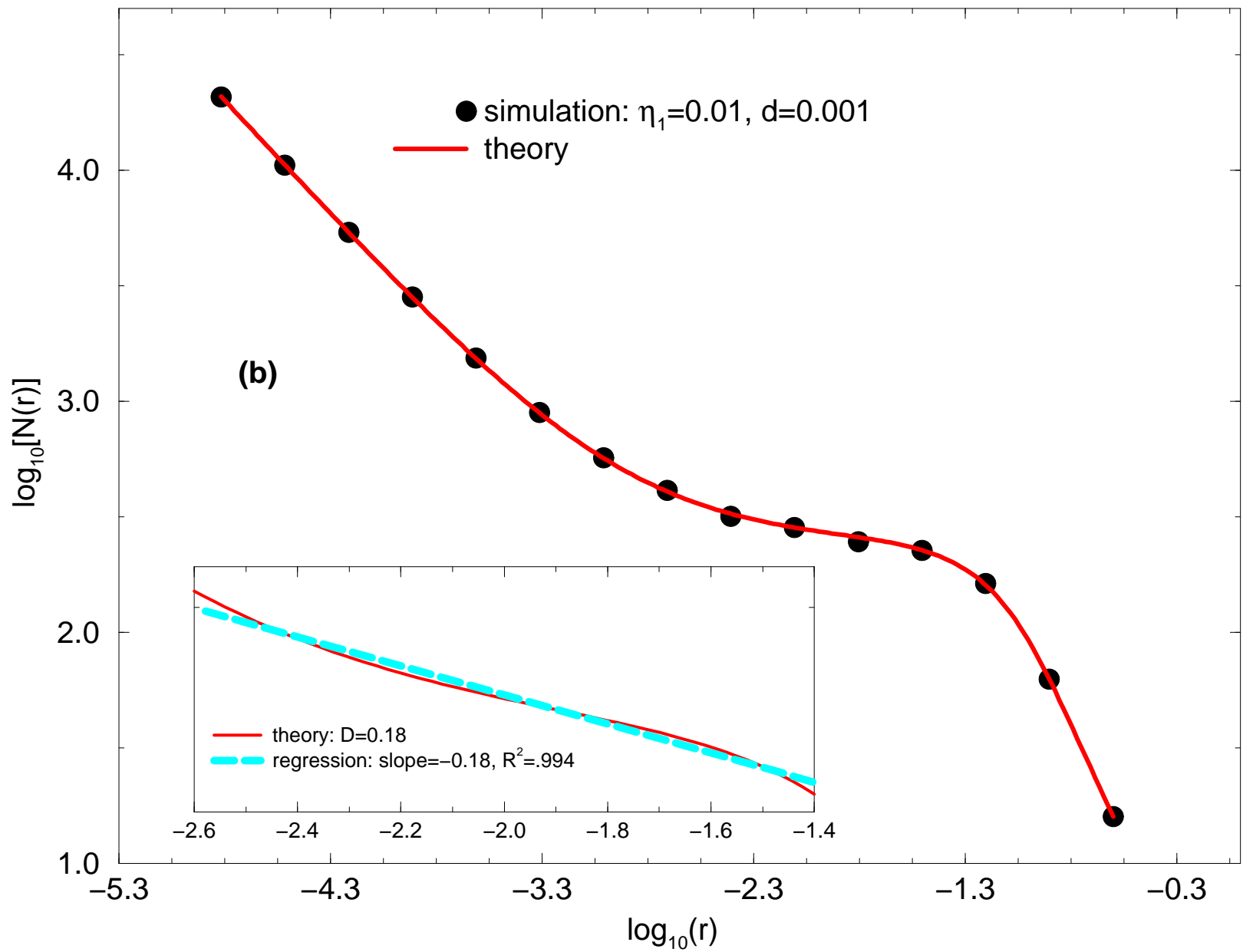
\end{document}